\documentclass[12pt]{article}
\usepackage{amssymb,amsmath}
\parskip=\medskipamount
\newtheorem{definition}{Definition}[section]

\newtheorem{proposition}[definition]{Proposition}

 1   
\font\ddpp=msbm10  scaled \magstep 1  

\def\cal#1{\mathcal{#1}}
\def\QED{\hskip0.1em\hfill\null\ \null\nobreak\hfill
\kern3pt\lower1.8pt\vbox{\hrule\hbox
{\vrule\kern1pt\vbox{\kern1.7pt \hbox{$\scriptstyle
QED$}\kern0.2pt}\kern1pt\vrule}\hrule}}
\def\R{\hbox{\ddpp R}}               

\def\hfl#1#2{\smash{\mathop{\hbox to 12 mm{\rightarrowfill}}
\limits^{\scriptstyle#1}_{\scriptstyle#2}}}

\begin{document}

\title{Skinner-Rusk approach to time-dependent mechanics}

\author{Jorge Cort\'es \\ Systems, Signals and Control Department,
Faculty of Mathematical Sciences\\ University of Twente, PO Box
217, 7500 AE Enschede, The Netherlands \\
e-mail: j.cortesmonforte@math.utwente.nl\\
Sonia Mart{\'\i}nez\\
Instituto de Matem\'aticas y F\'{\i}sicas
Fundamental\\ Consejo Superior de Investigaciones Cient\'{\i}ficas\\
Serrano 123, 28006~Madrid, Spain\\
e-mail: s.martinez@imaff.cfmac.csic.es\\
Frans Cantrijn\\
Department of Mathematical Physics and
Astronomy\\
Ghent University, Krijgslaan 281-S9, 9000 Ghent, Belgium\\
e-mail: frans.cantrijn@rug.ac.be}

\maketitle
\begin{abstract}
The geometric approach to autonomous classical mechanical systems
in terms of a canonical first-order system on the Whitney sum of
the tangent and cotangent bundle, developed by R. Skinner and R.
Rusk, is extended to the time-dependent framework.
\end{abstract}
\section{Introduction}
In 1983, it was shown by R. Skinner and R. Rusk that the dynamics
of an autonomous classical mechanical system, with configuration
space $Q$, can be properly represented by a first-order system on
the Whitney sum $T^{\ast}Q \oplus TQ$ \cite{SkRu1,SkRu2,SkRu3}. If
the system under consideration admits a Lagrangian description,
with Lagrangian $L \in C^{\infty}(TQ)$, the corresponding
first-order system on $T^{\ast}Q \oplus TQ$ is a Hamiltonian
system with respect to a canonical presymplectic structure. The
Skinner-Rusk formulation can be briefly summarized as follows.
Denoting the projections of $T^{\ast}Q \oplus TQ$ onto
$T^{\ast}Q$, resp.\ $TQ$, by $pr_1$, resp.\ $pr_2$, and putting
$\omega = pr_1^{\ast}\omega_Q$, with $\omega_Q$ the canonical
symplectic form on $T^{\ast}Q$, one can consider the following
equation
\begin{equation}\label{Skinner}
i_Z \omega = d {\cal H},
\end{equation}
where ${\cal H}:= \langle pr_1,pr_2 \rangle - L \circ pr_2$, and
$\langle \;,\; \rangle$ denotes the natural pairing between the
dual bundles $T^{\ast}Q$ and $TQ$. If the given Lagrangian $L$ is
regular, analysis of (\ref{Skinner}) shows that there exists a
unique solution $Z$ which is tangent to the graph of the Legendre
map $Leg_L: TQ \rightarrow T^{\ast}Q, (q^A,v^A) \mapsto (q^A,
{\partial L}/{\partial v^A})$, where the $q^A$ are local
coordinates on $Q$ and $(q^A,v^A)$ denote the corresponding bundle
coordinates on $TQ$.

{\bf Remark.} Here and in the sequel we will adopt the following
definition for the {\it graph of a bundle mapping}. Let $E_1$ and
$E_2$ denote two fibre bundles over the same base space $M$, and
let $f: E_1 \rightarrow E_2\;, (m,e) \mapsto (m,\tilde{f}_m(e))$
be a fibre bundle mapping over the identity. Then we define the
graph of $f$ as the image of the mapping $f \times_M id_{E_1}: E_1
\rightarrow E_2 \times_M E_1\;, (m,e) \in E_1 \mapsto
(m,\tilde{f}_m(e),e)$.

The equations of motion induced by the vector field $Z$ are
equivalent to the Euler-Lagrange equations for the system under
consideration (see \cite{SkRu2}). The important point now is that
this equivalence between a Lagrangian system and the corresponding
first-order system (\ref{Skinner}) on $T^{\ast}Q \oplus TQ$ also
holds if the given Lagrangian $L$ is singular. In that case, in
order to extract a consistent system of differential equations
from (\ref{Skinner}), one will have to invoke a constraint
algorithm. In fact, one of the main motivations for the work of
Skinner and Rusk was precisely to show that the Dirac-Bergmann
approach to singular Lagrangian systems can be properly described
on the Whitney sum of the tangent and cotangent bundle of the
configuration space, thereby avoiding some ambiguities occurring
in the literature on the subject.

The Skinner-Rusk formalism has been applied by several authors in
various contexts~\cite{CaLoRa,Lo,IbMa,CoLeMaMa}. As one of the
benefits of the formalism it turns out that it provides an
appropriate setting for a geometric approach to constrained
variational optimization problems. The latter are frequently
encountered, for instance, in mathematical economics and in
engineering. This has been demonstrated, in particular, for some
optimal control problems in~\cite{IbMa}, and for the so-called
vakonomic dynamics in~\cite{CoLeMaMa} (see also~\cite{CoMa}). The
fact that it would be interesting to extend those results to the
time-dependent framework, allowing for systems with an explicit
time-dependence of the `forces' and/or the constraints, is the
main motivation underlying the present work. More precisely, we
will develop here a time-dependent version of the Skinner-Rusk
formulation of dynamics, using the language of jet bundle
theory~\cite{Sa} and cosymplectic geometry~\cite{LeRo}. Among the
virtues of this new formulation of time-dependent mechanics, we
would like to stress the possibility it offers to model a large
class of systems, also in areas such as economics and control
theory. Applications for which this framework seems to be
particularly well-suited include stabilization and trajectory
tracking of mechanical systems by means of time-dependent
transformations (see e.g.\ \cite{FuSu}).

Our starting point is a fibre bundle  $\pi: E \longrightarrow \R$,
with $E$ representing the evolution space of a mechanical system.
Although it is quite common in treatments of time-dependent
mechanics to fix a trivialization of $\pi$, i.e.\ to work on a
direct product space $\R \times Q (= E)$, we will not select such
a trivialization here. The natural space to consider then for the
treatment of time-dependent Lagrangian mechanics is the first jet
space $J^1 \pi$, with the Lagrangian of the system being given as
a function $L \in C^{\infty}(J^1 \pi)$. The immediate candidate
for replacing the direct sum $T^{\ast}Q \oplus TQ$ in the
Skinner-Rusk model for autonomous systems, seems to be the fibred
product $J^1\pi^{\ast} \times_{E} J^1\pi$, where $J^1\pi^{\ast}$
is the `dual' of the affine bundle $J^1\pi$ (for this notion of
dual, see e.g.\ \cite{SaCaSa}). It turns out, however, that this
is not an appropriate choice for the following reasons. First,
there does not exist a natural pairing between $J^1\pi$ and
$J^1\pi^{\ast}$, needed for the construction of the time-dependent
analogue of the ``Hamiltonian" ${\cal H}$ appearing in
(\ref{Skinner}) and, secondly, there is no canonical 2-form on
$J^1\pi^{\ast}$ to take over the role of the symplectic form
$\omega_Q$ in the autonomous picture. To overcome these
difficulties we will show that the appropriate space to consider
is the fibred product $T^{\ast}E \times_E J^1\pi$.

Two final remarks are in order here. First of all, although we
will restrict ourselves to Lagrangian systems, it is clear that,
in analogy with the autonomous case (cf. \cite{SkRu1}), the
treatment can be easily extended to more general time-dependent
mechanical systems, with forces not necessarily derivable from a
potential. Secondly, it is interesting to note that the ideas
developed here also admit a further extension to classical field
theory, as has been demonstrated in a recent paper by M. de Le\'on
{\it et al.}~\cite{MaLeMar}.

The present paper is organized as follows. In the next section we
briefly recall the jet bundle approach to time-dependent
Lagrangian and Hamiltonian mechanics, including the description of
the constraint algorithm in case of a singular Lagrangian. In
Section 3 we then develop the Skinner-Rusk formalism for
time-dependent systems. We end with some conclusions and an
outlook on future work along these lines.

\section{Non-autonomous Lagrangian systems}

\subsection{The regular case}
Let $\pi: E \longrightarrow \R$ be a fibre bundle (the evolution
space), with $\hbox{dim}\,E = n+1$ and local bundle coordinates
$(t,q^A), A=1,\ldots,n$. Consider the corresponding 1-jet space
$J^1 \pi$, with coordinates $(t,q^A,\dot{q}^A)$ and associated
projections $\pi_1: J^1 \pi \longrightarrow \R$ and $\pi_{1,0}:
J^1 \pi \longrightarrow E$. Given a time-dependent Lagrangian $L:
J^1 \pi \longrightarrow \R$, the Euler-Lagrange equations read in
local coordinates
\begin{equation}\label{EL}
\frac{d}{dt} \left( \frac{\partial L}{\partial \dot{q}^A}\right) -
\frac{\partial L}{\partial q^A} = 0 \, .
\end{equation}
These equations can be rewritten in geometrical terms as follows.
Define the Poincar\'e-Cartan 1-form and 2-form
\[
\Theta_L=L \tilde{\eta} + \tilde{S}^{\ast}(dL) \, , \qquad
\omega_L=-d\Theta_L \, ,
\]
where $\tilde{\eta} = \pi_1^{\ast}(dt)$ and
$\displaystyle{\tilde{S}=
(dq^A-\dot{q}^Adt)\otimes\frac{\partial}{\partial \dot{q}^A}}$ is
the canonical vertical endomorphism on $J^1 \pi$ (see \cite{Sa}).
The action of $\tilde{S}$ on 1-forms is denoted by
$\tilde{S}^{\ast}$. The equations (\ref{EL}) can then be expressed
as
\begin{equation}\label{GEO}
i_{X}\omega_L = 0\,, \qquad i_{X}\tilde{\eta} = 1 \, .
\end{equation}

If the given Lagrangian is regular, then $\omega_L$ has maximal
rank and the pair $(\omega_L,\tilde{\eta})$ determines a {\it
cosymplectic structure} on $J^1\pi$, i.e.\ both forms are closed
and satisfy the conditions $\omega_L^n \wedge \tilde{\eta} \neq
0$, $\omega_L^{n+1}=0$ (cf.\ \cite{LeRo}). It then follows that
(\ref{GEO}) admits a unique solution, called the {\it
Euler-Lagrange vector field} for $L$, and which we will denote by
$X_L$. It is a second-order vector field, i.e.\ $\tilde{S}(X_L)=0$
and $i_{X_L}\tilde{\eta}=1$, and a direct computation shows that
integral curves of $X_L$ determine solutions of the Euler-Lagrange
equations (\ref{EL}) and vice-versa.

There also exists an alternative Hamiltonian description of the
problem. Consider the Legendre map $Leg_L: J^1 \pi \longrightarrow
T^{\ast}E$, defined by $Leg_L(j^1_t\phi)(v) =
(\Theta_L)_{j^1_t\phi} (\tilde{v})$ for $j^1_t\phi \in J^1 \pi$,
$v \in T_{\phi(t)}E$ and for any $\tilde{v} \in T_{j^1_t\phi} J^1
\pi$ such that ${\pi_{1,0}}_{\ast}(\tilde{v})=v$. Let $V\pi$
denote the subbundle of $TE$ consisting of $\pi$-vertical tangent
vectors, and denote its annihilator in $T^{\ast}E$ by $(V\pi)^o$.
Consider the quotient bundle $J^1 \pi^{\ast} =
T^{\ast}E/(V\pi)^o$, which is called the dual of $J^1\pi$, with
associated projections $\nu: T^{\ast}E \longrightarrow J^1
\pi^{\ast}$, $\tilde{\pi}_{1,0}: J^1 \pi^{\ast} \longrightarrow E$
and $\tilde{\pi}_1: J^1\pi^{\ast} \longrightarrow \R$. Finally,
denote by $leg_L:J^1 \pi \longrightarrow J^1 \pi^{\ast}$ the
composition $leg_L=\nu \circ Leg_L$. If $L$ is regular, then
$Leg_L$ is an immersion and $leg_L$ is a local diffeomorphism.

Assume now that the Lagrangian $L$ is hyperregular, that is,
$leg_L$ is a global diffeomorphism. Consider then the map $h=Leg_L
\circ leg_L^{-1}$. The mapping $h: J^1 \pi^{\ast} \longrightarrow
T^{\ast}E$ is a section of the projection $\nu$, i.e.\ $\nu \circ
h = id_{{}_{J^1\pi^{\ast}}}$ and is called a Hamiltonian of the
system. Next, denote by $\omega_E$ the canonical symplectic
two-form on $T^{\ast}E$ and let $\omega_h = h^{\ast} \omega_E$ be
its pull-back to $J^1 \pi^{\ast}$ under $h$. If $\eta_1 :=
(\tilde{\pi}_1)^{\ast}dt$, then $(\omega_h,\eta_1)$ defines a
cosymplectic structure on $J^1 \pi^{\ast}$. In addition, one has
that $leg_L^{\ast} (\omega_h) = \omega_L$ and $leg_L^{\ast} \eta_1
= \tilde{\eta}$. It then easily follows that the solution $X$ of
(\ref{GEO}) is $leg_L$-related to the (unique) solution $Y$ of the
equations
\begin{equation}\label{ham}
i_{Y}\omega_h = 0 \, , \qquad i_{Y}\eta_1 = 1 \, .
\end{equation}
Note that, always under the assumption of (hyper-)regularity of
$L$, the vector fields $X_L$, resp.\ $Y$, are precisely the Reeb
vector fields corresponding to the cosymplectic structures
$(\omega_L,\tilde{\eta})$, resp.\ $(\omega_h,\eta_1)$.

\subsection{The singular case: the constraint algorithm}

Suppose now that the given Lagrangian $L$ is degenerate, in the
sense that the Hessian matrix $({\partial^2L}/{\partial
\dot{q}^A}{\partial \dot{q}^B})$ is singular. We confine ourselves
to the case where this matrix has constant rank everywhere, say
$r$. The pair $(\omega_L,\tilde{\eta})$ then satisfies the
following relations (cf.\ \cite{ChLeMa,LeMaMa}):
\[
\omega_L^r \wedge \tilde{\eta} \neq 0;,\quad \omega_L^{r+1}\wedge
\tilde{\eta} = 0\;, \quad \omega_L^{r+2}=0.
\]
It follows that $2r \leq \hbox{rank}\;\omega_L \leq 2r+2$. In
general, the equations (\ref{GEO}) will not admit a global
solution $X$. Moreover, if a solution exists it will not be
unique. Therefore, in order to determine a consistent dynamics for
such a system (if it exists) one has to apply a constraint
algorithm which, at least in the case of Lagrangian mechanics,
should be supplemented with the ``second-order equation
condition". For completeness, we will now briefly sketch the
constraint algorithm described in~\cite{ChLeMa,LeMaMa}, which is
an adaptation to the time-dependent setting of the well-known
geometric constraint algorithm for presymplectic systems developed
by M. Gotay and J. Nester~\cite{GoNe1,GoNe2}.

With a view on its application later on, we will describe the
constraint algorithm here in the general framework of a structure
$(M,\Omega,\eta)$ consisting of a smooth manifold $M$, a closed
2-form $\Omega$ and a closed 1-form $\eta$, satisfying
\[
\Omega^r \wedge \eta \neq 0\;, \quad \Omega^{r+1}\wedge \eta =
0\;, \quad \Omega^{r+2}=0\;,
\]
for some $r < \dim M$. On $M$ we then consider the equations
\begin{equation}\label{preco}
i_X\Omega=0\;, \qquad i_X\eta=1\;.
\end{equation}
One can prove that there exists a vector $X_x \in T_xM$ satisfying
these equations at the point $x$ iff $\hbox{rank}\;\Omega_x = 2r$
(see \cite{ChLeMa}). In particular, it follows that (\ref{preco})
admits a global (but not necessarily unique) solution $X$ iff
$\Omega$ has constant rank $2r$, in which case the given pair
$(\Omega,\eta)$ defines a so-called {\it precosymplectic
structure} on $M$. If this is not the case, the constraint
algorithm proceeds as follows. Put $P_1:=M$ and consider the set
\[
P_2:=\{x \in M\;|\; \exists \, X_x \in T_x M \;\hbox{such that}\;
i_{X_x} \Omega_x = 0\,, \; i_{X_x} \eta_x = 1 \}\;.
\]
According to the previous observation, this set can be
equivalently characterized by $P_2 = \{x \in P_1\;|\;
\hbox{rank}\;\Omega_x = 2r\}$. We then assume that $P_2$ is an
embedded submanifold of $P_1(=M)$ and we denote the natural
inclusion by $j_2: P_2 \hookrightarrow P_1$. We are then assured
that the equations (\ref{preco}) admit a solution $X$ defined at
all points of $P_2$, but $X$ need not be tangent to $P_2$ and,
hence, does not necessarily induce a dynamics on $P_2$. We
therefore have to continue the constraint algorithm by considering
the subset
\begin{align*}
P_3:&= \{x \in P_2\;|\;\exists \, X_x \in T_x P_2 \;\hbox{such
that}\; i_{X_x} \Omega(x) = 0\,, \; i_{X_x} \eta (x) = 1 \}\\
&=\{x \in P_2\;|\; \eta_x \in \flat(T_xP_2)\}\;,
\end{align*}
where $\flat$ is the bundle morphism defined by
\[
\flat: TM \longrightarrow T^{\ast}M\;, v \in T_xM \longmapsto
i_v\Omega_x + (i_v\eta_x)\eta_x\;.
\]
Assuming $P_3$ is a submanifold of $P_2$, with inclusion map $j_3:
P_2 \hookrightarrow P_1$, it follows that there exists a vector
field $X$ on $P_2$, which satisfies (\ref{preco}) at points of
$P_3$. Again, however, such an $X$ need not be tangent to $P_3$,
and one may have to repeat the above procedure. In this way, a
descending sequence of submanifolds
\[
\ldots \stackrel{j_{\ell+1}}{\hookrightarrow} P_{\ell}
\stackrel{j_{\ell}}{\hookrightarrow} \ldots
\stackrel{j_4}{\hookrightarrow} P_3
\stackrel{j_3}{\hookrightarrow} P_2
\stackrel{j_2}{\hookrightarrow} P_1(=M)
\]
is generated, with
\begin{equation}\label{M}
P_{\ell} := \{x \in P_{\ell-1}\;|\; \eta_x \in
\flat(T_xP_{\ell-1}\}\quad (\ell \geq 2)\;,
\end{equation}
and where $P_{\ell}$ is called the {\it $\ell$-ary constraint
submanifold}. If this sequence terminates at a nonempty set, in
the sense that for some finite $k \geq 1$ we have $P_{k+1}=
P_{k}$, but $P_k \neq P_{k-1}$, then $P_k$ is called the {\it
final constraint submanifold}, which we denote by $P_f$. Now, it
may still happen that $\dim P_f = 0$ (i.e.\ $P_f$ is a discrete
set), in which case the given problem admits no proper dynamics.
However, if $\dim P_f > 0$, then we know by construction that
there exists a vector field $X$ on $M$, defined along $P_f$, which
is tangent to $P_f$ and satisfies the equations
\[
i_X\Omega_{|P_f} = 0\;, \quad i_X\eta_{|P_f}=1\;,
\]
i.e.\ the given dynamical problem admits a consistent solution on
$P_f$. In general, however, this solution will not be unique:
given a solution $X$, for any smooth section $Y$ of the bundle
$(\hbox{ker}\;\Omega \cap \hbox{ker}\;\eta) \cap TP_f$ over $P_f$,
$X + Y$ is also a solution.

If we are dealing with a time-dependent Lagrangian system, i.e.\
with $M=J^1\pi, \Omega = \omega_L, \eta=\tilde{\eta}\,(= dt)$,
this is not the full story. First of all, we then also have to
impose the so-called ``second-order differential equation"
condition, i.e.\ we are only interested in a solution $X$ which
determines a system of second-order ordinary differential
equations. Secondly, as in the autonomous case, one can develop a
similar constraint algorithm on the Hamiltonian side (i.e.\ on
$J^1\pi^{\ast}$) and, under a suitable assumption regarding the
given Lagrangian, one can show that both descriptions are
equivalent. For more details, we again refer to
\cite{ChLeMa,LeMaMa}.

In the next section we will show how the above can be translated
into a Skinner-Rusk type formulation for time-dependent Lagrangian
systems.

\section{Skinner-Rusk formulation}

We start again from a fibre bundle $\pi: E \rightarrow \R$, with
$n$-dimensional fibre, and its first jet space $J^1\pi$. Bundle
coordinates on $E$ and $J^1\pi$ are denoted by $(t,q^A)$ and
$(t,q^A,\dot{q}^A)$, respectively. Canonical coordinates on
$T^{\ast}E$ will be written as $(t,q^A,\tau,p_A)$ and the
canonical symplectic form on $T^{\ast}E$ then reads $\omega_E =
dq^A \wedge dp_A + dt \wedge d\tau$. We now consider the fibred
product of the bundles $T^{\ast}E$ and $J^1 \pi$ over $E$, i.e.\
$T^{\ast}E \times_E J^1\pi$, with projections $pr_1: T^{\ast}E
\times_E J^1\pi \rightarrow T^{\ast}E$, $pr_2:T^{\ast}E \times_E
J^1\pi \rightarrow J^1\pi$ and $pr: T^{\ast}E \times_E J^1 \pi
\rightarrow E$. The natural bundle coordinates on $T^{\ast}E
\times_E J^1\pi$ are $(t,q^A,\tau,p_A,\dot{q}^A)$.

On $T^{\ast}E \times_E J^1 \pi$ we define the $2$-form $\omega$ as
the pullback of the canonical symplectic form on $T^{\ast}E$,
i.e.\ $\omega=pr_1^{\ast} \omega_E$, and the $1$-form $\eta = (\pi
\circ pr)^{\ast}(dt) = pr_2^{\ast}\tilde{\eta}$. For simplicity we
will sometimes write $\eta = dt$. Recall that the affine bundle
$J^1\pi$ can be identified with an affine subbundle of $TE$ whose
underlying set is given by $\{v\in TE\:|\; \langle dt,v \rangle =
1\}$. In coordinates, the natural embedding $j: J^1\pi
\hookrightarrow TE$ reads $j(t,q^A,\dot{q}^A) =
(t,q^A,1,\dot{q}^A)$.

Given a Lagrangian $L \in C^{\infty}(J^1{\pi})$, we can define the
following function on $T^{\ast}E \times_E J^1\pi$:
\[
{\cal H} = \langle pr_1,j \circ pr_2 \rangle - pr_2^{\ast} L \, ,
\]
where $\langle \;,\; \rangle$ denotes the natural pairing between
vectors and covectors on $E$. In coordinates this becomes ${\cal
H} = p_A \dot{q}^ A + \tau - L(t,q^A,\dot{q}^A)$. Putting
\[
\omega_{\cal H} = \omega + d{\cal H} \wedge \eta\;,
\]
we can then consider the following equations:
\begin{equation}\label{SR}
i_{Z}\omega_{\cal H} = 0 \, , \qquad i_{Z}\eta = 1 \, .
\end{equation}
Let us try to find out, in coordinates, what kind of dynamics is
encoded by (\ref{SR}). For that purpose, we write $Z$ as
\[
Z = Z_t \frac{\partial }{\partial t} + Z_{q^A} \frac{\partial
}{\partial q^A}+ Z_\tau \frac{\partial }{\partial \tau} + Z_{p_A}
\frac{\partial }{\partial p_A} + Z_{\dot{q}^A} \frac{\partial
}{\partial \dot{q}^A}\, .
\]
From the second equation in (\ref{SR}) we deduce that $Z_t = 1$,
and the first equation then becomes:
\begin{align*}
i_{Z}\omega_{\cal H} &= i_{Z}\omega + Z({\cal H})dt - d{\cal H} \\
&= (Z({\cal H})+ \frac{\partial L}{\partial t} -Z_\tau )dt +
(\frac{\partial L}{\partial q^A} - Z_{p_A}) dq^A - (p_A -
\frac{\partial L}{\partial \dot{q}^A})d\dot{q}^A + (Z_{q^A} -
\dot{q}^A) dp_A \\
&= 0 \, .
\end{align*}
This immediately gives $Z_{q^A} = \dot{q}^A$ and $\displaystyle{
Z_{p_A} = \frac{\partial L}{\partial q^A}}$, together with the
following constraint equations: $\displaystyle{p_A=\frac{\partial
L}{\partial \dot{q}^A}}$. These constraints determine a
submanifold of $T^{\ast}E \times_E J^1\pi$ which, for convenience,
we will denote by $M_L$. With the above expressions for
$Z_{q^A},Z_{p_A}$ and $Z_{t}$, we see that the remaining condition
$\displaystyle{ Z({\cal H})+ \frac{\partial L}{\partial t} -Z_\tau
= 0 }$ is identically satisfied at all points of $M_L$,
irrespective of the value of the components $Z_{\tau}$ and
$Z_{\dot{q}^A}$. Note that the relation $Z_{q^A}=\dot{q}^A$
reflects the second-order differential equation property.

The previous analysis already shows that (\ref{SR}) locally admits
a solution $Z$, defined in points of the submanifold $M_L$ of
$T^{\ast}E \times_E J^1\pi$. In fact, we have a whole family of
solutions since the components $Z_{\tau}$ and $Z_{\dot{q}^A}$ can
still be chosen arbitrarily.

In order to obtain consistent equations of motion, however, we
have to impose the condition that $Z$ be tangent to the
submanifold $M_L$, that is, the functions $\displaystyle{ Z(p_A -
\frac{\partial L}{\partial \dot{q}^A})}$ should vanish at points
of $M_L$ for all $A=1, \ldots, n$. We now have that
\begin{eqnarray}\label{tan}
Z(p_A - \frac{\partial L}{\partial \dot{q}^A}) = \frac{\partial
L}{\partial q^A} - \frac{\partial^2 L}{\partial t\partial
\dot{q}^A} - \dot{q}^B \frac{\partial^2 L}{\partial q^B \partial
\dot{q}^A}  - Z_{\dot{q}^B} \frac{\partial^2 L}{\partial \dot{q}^A
\partial \dot{q}^B} \, .
\end{eqnarray}
Clearly, if $L$ is regular, the vanishing of (\ref{tan}) fixes all
the components $Z_{\dot{q}^A}$ as functions of $(t,q^A,\dot{q}^A)$
on $M_L$. If not, one will have to apply a constraint algorithm.

\subsection{The regular case}
Let us assume that the given Lagrangian $L$ is regular. The
previous analysis tells us that the system (\ref{SR}) admits a
solution $Z$ on $M_L$ and it follows from the expression for the
components $Z_t,Z_{q^A},Z_{p_A},Z_{\dot{q}^A}$ that any integral
curve of $Z$ on $M_L$ determines a solution $(q^A(t))$ of the
corresponding Euler-Lagrange equations of motion
\[
\frac{d}{dt} \left( \frac{\partial L}{\partial \dot{q}^A}\right) -
\frac{\partial L}{\partial q^A} = 0 \,, \qquad (A=1,\ldots,n)\,.
\]
However, the solution $Z$ is not unique since the component
$Z_{\tau}$ is still undetermined. This, of course, is not
surprising since $\partial/{\partial \tau}$ belongs to
$\hbox{ker}\;\omega_{\cal H} \cap \hbox{ker}\;\eta$, i.e.
$i_{\frac{\partial}{\partial \tau}}\omega_{\cal H} = 0$ and
$i_{\frac{\partial}{\partial \tau}}\eta =0$. In order to obtain a
unique dynamics, we now impose an additional constraint
\[
\tau = L - \dot{q}^A\frac{\partial L}{\partial \dot{q}^A}\,.
\]
Together with the constraints $\displaystyle{p_A=\frac{\partial
L}{\partial \dot{q}^A}}$, these are (locally) the defining
equations of a submanifold of $T^{\ast}E \times_E J^1\pi$, namely
the graph of the (extended) Legendre map
\[
Leg_L: J^1\pi \longrightarrow T^{\ast}E\,,(t,q^A,\dot{q}^A)
\longmapsto (t,q^A,L - \dot{q}^A\frac{\partial L}{\partial
\dot{q}^A},\frac{\partial L}{\partial \dot{q}^A})
\]
(for the intrinsic definition of $Leg_L$, see e.g.\ \cite{SaCaSa},
and for the notion of graph considered here, see the Remark in the
Introduction). We denote the graph of $Leg_L$ by $\hbox{graph}_L$.
Clearly, $\hbox{graph}_L \subset M_L$ and if we now require that
$Z$ should be tangent to $\hbox{graph}_L$, it readily follows that
\begin{equation}\label{tau}
Z_{\tau} = Z(L - \dot{q}^A\frac{\partial L}{\partial
\dot{q}^A})\,,
\end{equation}
which uniquely fixes $Z_{\tau}$. Note that the differential
equation corresponding to the $\tau$-component of $Z$ represents
the so-called ``energy-balance" equation from time-dependent
mechanics.

The above construction was carried out on an arbitrary natural
bundle chart of $T^{\ast}E \times_E J^1\pi$. Using a standard
argument it then follows that $Z$ is in fact well-defined on the
whole of $\hbox{graph}_L$. Defining the mapping
\[
Leg_L\times_E id_{J^1\pi}: J^1\pi \longrightarrow T^{\ast}E
\times_E J^1\pi\;,\; (t,q^A,\dot{q}^A) \longmapsto (t,q^A,L -
\dot{q}^A\frac{\partial L}{\partial \dot{q}^A},\frac{\partial
L}{\partial \dot{q}^A},\dot{q}^A)\;,
\]
we see that $\hbox{Im}(Leg_L\times_E id_{J^1\pi})= \hbox{graph}_L$
and it is not difficult to verify that the unique solution $Z$ of
(\ref{SR}), defined on $\hbox{graph}_L$, and the Euler-Lagrange
vector field $X_L$ on $J^1\pi$ are related by
\[
(Leg_L\times_E id_{J^1\pi})_{\ast}(X_L)_x = Z_{\bar{x}}\;,
\]
where $\bar{x}= (Leg_L\times_E id_{J^1\pi})(x)$, for all $x \in
J^1\pi$.

The previous discussion can now be summarized by the following
proposition.

\begin{proposition} For a regular Lagrangian $L$, the system
(\ref{SR}) admits a unique solution $Z$ defined on, and tangent
to, $\hbox{graph}_L$, and the induced vector field on
$\hbox{graph}_L$ is $(Leg_L\times_E id_{J^1\pi})$-related to the
Euler-Lagrange vector field on $J^1\pi$.
\end{proposition}

Recall that there exists a canonical projection $\nu: T^{\ast}E
\rightarrow J^1\pi^{\ast}\,(= T^{\ast}E/(V\pi)^0)$ which, in
coordinates, reads $\nu(t,q^A,\tau,p_A) = (t,q^A,p_A)$ (cf.
Section 2.1 and \cite{SaCaSa}). Let us assume that $L$ is
hyperregular such that the mapping $leg_L: = \nu \circ Leg_L$ is a
global diffeomorphism.  Consider the fibred product $J^1
\pi^{\ast} \times_E J^1 \pi$, with associated projections
$\lambda_1$ and $\lambda_2$ onto $J^1\pi^{\ast}$ and $J^1{\pi}$,
respectively. We can then define the following projection
\[
\nu \times_E id_{J^1\pi} : T^*E \times_E J^1 \pi \longrightarrow
J^1 \pi^* \times_E J^1 \pi\,, (t,q^A, \tau, p_A, \dot{q}^A)
\longmapsto (t,q^A, p_A, \dot{q}^A)\,.
\]
From the discussion above we deduce that, along $\hbox{graph}_L$,
the vertical distribution determined by the projection $\nu
\times_E id_{J^1\pi}$, i.e.\ $\hbox{ker} (\nu \times_E id_{J^1\pi}
)_{\ast}$, is invariant under $Z$ in the sense that
\[
[Z, \hbox{ker} (\nu \times_E id_{J^1\pi} )_{\ast}] \subset
\hbox{ker} (\nu \times_E id_{J^1\pi})_{\ast}\,.
\]
Hence, in the hyperregular case, the solution vector field $Z$ is
projectable onto $J^1\pi^{\ast}\times_E J^1\pi$. Its projection
locally reads
\[
(\nu \times_E id_{J^1\pi})_{\ast} (Z) = \frac{\partial }{\partial
t} + \dot{q}^A \frac{\partial }{\partial q^A} + Z_{\dot{q}^A}
(t,q^A,\dot{q}^A) \frac{\partial }{\partial \dot{q}^A} +
\frac{\partial L}{\partial q^A}(t,q^A,\dot{q}^A) \frac{\partial
}{\partial p_A} \, .
\]
This is the unique vector field on $J^1 \pi^{\ast} \times_E J^1
\pi$ determined by the equations
\[
i_{\tilde{Z}} (\lambda_1^{\ast} \omega_h) = 0 \, , \quad
i_{\tilde{Z}} (\lambda_1^{\ast} \eta_1) = 1 \, ,
\]
where we recall that $h = Leg_L \circ leg_L^{-1}$ and $\omega_h =
h^{\ast}\omega_E$ (cf.\ Section 2.1).

In the general case, however, it is not possible to represent the
dynamics of the non-autonomous problem corresponding to $L$ by a
first-order system on $J^1 \pi^* \times_E J^1 \pi $.

\subsection{The singular case}

Returning to the beginning of this section, let us now assume that
$L$ is not regular. Observe that, with $\omega :=
pr_1^{\ast}\omega_E$, we have
\[
\omega_{\cal H}^2 (= \omega_{\cal H} \wedge \omega_{\cal H}) =
\omega \wedge \omega + 2 \, \omega \wedge d{\cal H} \wedge \eta
\,,
\]
and, in general,
\[
\omega_{\cal H}^k = \omega^k + k \, \omega^{k-1} \wedge d{\cal H}
\wedge \eta \,,\quad \forall k\,.
\]
Herewith, it is straightforward to check that the pair
$(\omega_{\cal H}, \eta)$ satisfies the following relations:
\[
\omega_{\cal H}^n \wedge \eta \neq 0 \,, \;\; \omega_{\cal
H}^{n+1} \wedge \eta =0 \,,\;\; \omega_{\cal H}^{n+2}=0\,.
\]
Indeed, we have
\begin{align*}
\omega_{\cal H}^n \wedge \eta &= \omega^n \wedge \eta =
(-1)^{\frac{n(n+1)}{2}}n!\, dq^1 \wedge \dots \wedge dq^n
\wedge dp_1 \wedge \dots \wedge dp_n \wedge \eta \neq 0 \,, \\
\omega_{\cal H}^{n+1} \wedge \eta &=
(-1)^{\frac{(n+1)(n+2)}{2}}(n+1)! \,dt \wedge dq^1 \wedge \ldots
\wedge dq^n \wedge d\tau \wedge dp_1 \wedge
\ldots \wedge dp_n \wedge \eta = 0 \,,\; \\
\omega_{\cal H}^{n+2} &= \omega^{n+2} + (n+2)\, \omega^{n+1}
\wedge d{\cal H} \wedge dt = 0\,.
\end{align*}
This implies, in particular, that $2n \leq
\hbox{rank}\,\omega_{\cal H} \leq 2(n+1)$, where we recall that
$\hbox{dim}\,E = n + 1$.

Putting $M_1:= T^{\ast}E \times_E J^1\pi$, we can now apply the
constraint algorithm described in Section 2.2 to the triplet
$(M_1,\omega_{\cal H},\eta)$. First of all, we consider the set
\[
M_2 = \{ x \in M_1 \,|\, \exists \, Z_x \in T_x M_1 \;\hbox{such
that}\; i_{Z_x} \omega_{\cal H}(x) = 0\,, \; i_{Z_x} \eta(x) = 1
\, \}\,,
\]
which can be equivalently characterized by $M_2 = \{ x \in M_1
\,|\, \hbox{rank}\,\omega_{\cal H}(x) = 2n \}$. In local
coordinates we find
\begin{eqnarray*}
\omega_{\cal H}^{n+1} (x) &=& \omega^{n+1} (x)+ (n+1)\, \omega^n
\wedge d{\cal H} \wedge \eta (x) \\
&=&  \omega^{n+1} (x) - (n+1)\,\omega^{n} \wedge \frac{\partial
{\cal H}}{\partial \tau} d\tau \wedge \eta (x) + (n
+1)\frac{\partial {\cal H}}{\partial \dot{q}^A}\omega_E^n
\wedge d \dot{q}^A \wedge \eta (x)\\
&=&  (-1)^{\frac{n(n+1)}{2}} (n+1)! \frac{\partial {\cal
H}}{\partial \dot{q}^A} dq^1 \wedge \dots \wedge dq^n \wedge dp_1
\wedge \dots \wedge dp_n  \wedge d\dot{q}^A \wedge \eta (x)\,,
\end{eqnarray*}
such that $x \in M_2$ if and only if
\[
\frac{\partial {\cal H}}{\partial \dot{q}^A}_{|x} \equiv (p_A -
\frac{\partial L}{\partial \dot{q}^A})_{|x} = 0\,,\;\; A =1,\ldots
n\,.
\]
Observe that $M_2$ coincides with the submanifold $M_L$ introduced
at the beginning of this section. By construction, there exists a
vector field $Z$ on $M_1$, defined along $M_2$, which verifies
equations (\ref{SR}) at points of $M_2$. But in general $Z$ will
not be tangent to $M_2$ and so we then have to proceed with the
constraint algorithm by considering the set
\[
M_3 = \{ x \in M_2 \, | \, \exists \, Z_x \in T_x M_2 \;\hbox{such
that}\; i_{Z_x} \omega_{\cal H} (x) = 0\,, \; i_{Z_x} \eta (x) = 1
\}\,.
\]
Assuming that $M_3$ is a smooth submanifold, there will be a
vector field $Z$ defined along  $M_3$ and tangent to $M_2$,
satisfying (\ref{SR}) at each point of $M_3$. Continuing this way,
we obtain a descending sequence of submanifolds of $M_1$ that, in
the favorable case, will stop at a final constraint submanifold
$M_f$ on which there exists a consistent solution of the given
dynamical problem (cf.\ Section 2.2). The constraint submanifolds
$M_{\ell}$ can still be characterized in an algebraic way similar
to (\ref{M}), with the map $\flat: TM_1 \rightarrow T^{\ast}M_1$
being induced here by the pair $(\omega_{\cal H},\eta)$.

As in the autonomous case, we thus see that the constraint
algorithm for time-dependent singular Lagrangian systems can be
properly developed in terms of the structure $(T^{\ast}E \times_E
J^1\pi, \omega_{\cal H},\eta)$. To complete the picture, we have
the following result which shows that this description is
equivalent to the standard one based on the structure
$(J^1\pi,\omega_L,\tilde{\eta})$.

\begin{proposition}\label{equiv} Let $\{P_{\ell}\}_{\ell \geq 1}$, resp.
$\{M_{\ell}\}_{\ell \geq 1}$, denote the sequence of constraint
submanifolds generated by applying the constraint algorithm to
$(J^1\pi,\omega_L,\tilde{\eta})$, resp. $(T^{\ast}E \times_E
J^1\pi, \omega_{\cal H},\eta)$. Then, for each $i = 1, 2, \ldots$,
we have that $\varphi_{i+1}\equiv{pr_2}_{|M_{i+1}}: M_{i+1}
\rightarrow P_i$ is a surjective submersion such that the
following diagram commutes
\begin{center}
\unitlength=1mm \special{em:linewidth 0.4pt} \linethickness{0.4pt}
\begin{picture}(80,88)
\put(10,80){\makebox(0,0){$M_1:=T^{\ast}E \times_E J^1 \pi$}}
\put(10,60){\makebox(0,0){$M_2$}}
\put(10,40){\makebox(0,0){$M_3$}}
\put(70,60){\makebox(0,0){$P_1:=J^1\pi$}}
\put(70,40){\makebox(0,0){$P_2$}} \put(10,64){\vector(0,1){12}}
\put(10,44){\vector(0,1){12}} \put(70,44){\vector(0,1){12}}
\put(28,76){\vector(3,-1){30}} \put(20,60){\vector(1,0){36}}
\put(20,40){\vector(1,0){36}} \put(4,70){\makebox(0,0){$j'_2$}}
\put(4,50){\makebox(0,0){$j'_3$}}
\put(38,64){\makebox(0,0){$\varphi_2$}}
\put(38,44){\makebox(0,0){$\varphi_3$}}
\put(74,50){\makebox(0,0){$j_2$}}
\put(50,75){\makebox(0,0){$pr_2$}} \put(10,32){\makebox(0,0){.}}
\put(10,28){\makebox(0,0){.}} \put(10,24){\makebox(0,0){.}}
\put(70,32){\makebox(0,0){.}} \put(70,28){\makebox(0,0){.}}
\put(70,24){\makebox(0,0){.}}
\end{picture}
\end{center}
\vspace*{-1.7cm} (where $j_{\ell}: P_{\ell} \rightarrow P_{\ell
-1}$ and $j'_{\ell}: M_{\ell} \rightarrow M_{\ell - 1}$ are the
natural embeddings of the respective constraint submanifolds).
Moreover, if there is a final constraint submanifold $M_f:= M_k
\subset M_1$ (for some $k \geq 2$) on which there exists a
consistent solution $Z$ of (\ref{SR}), then $Z$ projects under
$\varphi_k$ onto a solution of (\ref{GEO}) on the final constraint
submanifold $P_f:=P_{k-1}$ in $J^1\pi$ and, conversely, any
solution of (\ref{GEO}), defined on $P_f$, is the projection of a
vector field on $M_f$ which is a solution of (\ref{SR}).
\end{proposition}
The proof of this proposition essentially relies on the following
two facts. First of all, the projection $pr_2: T^{\ast}E \times_E
J^1\pi \rightarrow J^1\pi$ has the appropriate ``almost
regularity" properties, that is: (i) $pr_2$ is a surjective
submersion and (ii) the fibres of this submersion are connected
submanifolds of $T^{\ast}E \times_E J^1\pi$, being diffeomorphic
to $\R^{n+1}$. And, secondly, a straightforward computation shows
that
\begin{align*}
(pr_2^{\ast}\omega_L)(x) &= \left(dq^A \wedge
d\left(  \frac{\partial L}{\partial \dot{q}^A }\right) + dE_L \wedge dt\right)(x) \\
&= \left(dq^A \wedge d\left(\frac{\partial L}{\partial \dot{q}^A
}\right) + \dot{q}^A d\left(\frac{\partial L}{\partial \dot{q}^A
}\right) \wedge dt - \left(\frac{\partial L}{\partial q^A
}\right)dq^A \wedge dt\right)(x)\\
&= \omega_{\cal H} (x)\,,
\end{align*}
and, clearly, we also have $pr_2^{\ast}\tilde{\eta} = \eta$.
Herewith, the proof of Proposition \ref{equiv} can be easily
completed, following the same reasoning as in the autonomous
case~\cite{SkRu2}.

The solution generated by the constraint algorithm (if it exists)
is not unique. On the other hand, we may observe that if the given
dynamical problem (\ref{SR}) admits a consistent solution $Z$ on a
final constraint submanifold $M_f$ then, by construction, its
projection onto $J^1\pi$ will automatically verify the
second-order equation condition along a submanifold of $P_f$. This
is again in full analogy with the situation encountered in the
autonomous case.

Next, assume that the given Lagrangian $L \in C^{\infty}(J^1\pi)$
is almost regular in the following sense: (i) $Leg_L(J^1\pi)$ is a
submanifold of $T^{\ast}E$, (ii) $Leg_L$, regarded as a map from
$J^1\pi$ onto its image, is a submersion with connected fibres,
(iii) $leg_L^{-1}(leg_L(x))$ is a connected set for all $x \in
J^1\pi$. In \cite{LeMaMa} it has been shown that, with these
assumptions, one can develop a constraint algorithm on
$J^1\pi^{\ast}$ which is equivalent to the one on $J^1\pi$. Again
as in the autonomous case (see \cite{SkRu2}), one can demonstrate
that a solution of the constrained analysis on $J^1\pi^{\ast}$ can
be related to a solution $Z$ of (\ref{SR}), defined on the final
constraint submanifold $M_f$. This connection is established by
choosing a suitable (local) section $\sigma$ of the projection
$\nu \circ pr_1: T^{\ast}E \times_E J^1\pi \rightarrow
J^1\pi^{\ast}$ and restricting $Z$ to $\hbox{Im}(\sigma) \cap M_f$
(recall that $\nu$ is the canonical projection of $T^{\ast}E$ onto
$J^1\pi^{\ast}$).

\section{Conclusions}

We have developed a non-autonomous version of the Skinner-Rusk
approach to (Lagrangian) mechanics and have shown that, both in
the regular and in the singular case, this yields a first-order
system on the fibred product $T^{\ast}E \times_E J^1{\pi}$ which
encodes all the information of the dynamics of the system under
consideration. This approach to time-dependent mechanics possesses
the same virtues as in the autonomous case, such as the fact that
the `Hamiltonian' $\cal H$ is defined without having to solve the
relations $p_A = {\partial L}/{\partial \dot{q}^A}$ for (some of)
the velocities.

Within the above framework for the description of time-dependent
mechanics, there are several lines of investigation that seem to
be worth pursuing,  such as: the role and the nature of gauge
transformations in the case of singular Lagrangians and the
general study of symmetries of (time-dependent) mechanical
systems. In addition, it would certainly be of interest to use
this formalism for establishing a geometric formulation of optimal
control problems with an explicit time-dependence, such as in the
case of time-dependent vakonomic dynamics, thereby generalizing
the work presented in~\cite{CoLeMaMa,CoMa}.

{\bf Acknowledgements}\\
This research has been partially supported by a FPI grant from the
Spanish MCYT,  by grant DGICYT  PGC2000-2191-E and by the European
Union through the Training and Mobility of Researchers Program ERB
FMRXCT-970137. FC also wishes to acknowledge support from a
research grant of the ``Bijzonder Onderzoeksfonds" of Ghent
University.

\end{document}